\documentclass[twocolumn]{revtex4-1}
\usepackage[utf8]{inputenc}
\usepackage{graphicx}
\usepackage{textcomp}
\usepackage{txfonts}
\usepackage{color}
\usepackage{gensymb} 

\renewcommand{\v}[1]{{\bf #1}}
\newcommand{\dublin}{School of Physics, AMBER and CRANN Institute, Trinity College, Dublin 2, Ireland}
\newcommand{\prague}{Institute of Physics, Academy of Sciences of the Czech Republic,
Na Slovance 2, 182 21 Prague 8, Czech Republic}

\newcommand{\new}[1]{\textcolor{black}{#1}}

\begin{document}

\title{Electron trapping by neutral pristine ferroelectric domain walls in BiFeO$_3$}

\author{Sabine K\"{o}rbel}
\affiliation{\dublin}
\affiliation{\prague}
\author{Jirka Hlinka}
\affiliation{\prague}
\author{Stefano Sanvito}
\affiliation{\dublin}

\begin{abstract}
  First-principles calculations for pristine neutral ferroelectric domain walls in BiFeO$_3$ reveal that excess 
  electrons are selectively trapped by the domain walls, while holes are only weakly attracted. Such trapped 
  excess electrons \new{may} be responsible for the thermally activated electrical conductivity at domain walls 
  observed in experiments. In the case of a periodic array of domain walls, the trapped excess electrons create
  a zigzag potential, whose amplitude depends on the electron concentration in the material and the
  domain-wall distance. The potential is asymmetric for 71\textdegree~and 109\textdegree~domain walls.
  This could modify the open-circuit voltage in a solar cell and hence influence the photoelectric 
  effect in BiFeO$_3$.
\end{abstract}
\maketitle
\begin{figure}[htb]
   \def\svgwidth{0.4\textwidth}
   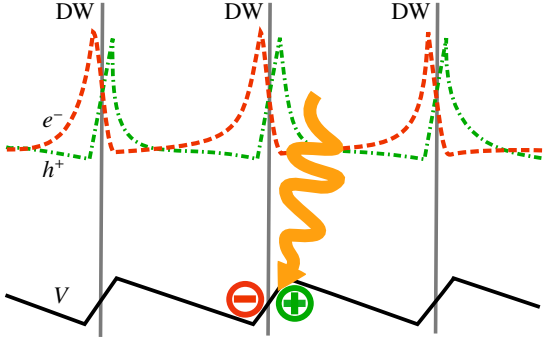
   \caption{\label{fig:model_seidel}(Color online) Previously proposed model for the charge-carrier 
   distribution (upper panel) of excess electrons $e^-$ (orange dashed line), holes $h^+$ 
   (green dot-dashed line), and the electronic potential (lower panel), $V$, at domain walls. All 
   curves are schematic.
  }
\end{figure}
\begin{figure}[b]
   \def\svgwidth{0.4\textwidth}
   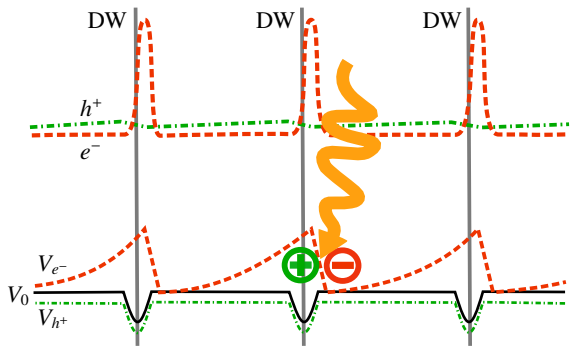
   \caption{\label{fig:model}(Color online) Our proposed model for charge-carrier distribution and 
   electronic potential. In the lower panel we show the electronic potential for the charge-neutral 
   system $V_0$ (black solid line), and in the case of excess electrons $V_{e^-}$ or holes $V_{h^+}$.
  }
\end{figure}

%
\paragraph*{Introduction.} 
%
The ferroelectric oxide BiFeO$_3$ exhibits the photovoltaic effect, which makes it a prototype material to study ferroelectric
photovoltaics.
In principle, ferroelectric photovoltaics are promising materials for solar cells because their polar structure
allows one to extract a photocurrent without the need to create a $p$-$n$ junction by doping.
In practice, ferroelectrics are usually large band-gap materials, with low light absorption in the visible range.
Two quantities are important for an efficient solar-cell absorber: the photocurrent, and the photovoltage, both of which should ideally be large.
Ferroelectrics can at least fulfill one of the two requirements, in that their open-circuit voltage can be very large.
This feature has first been ascribed to ferroelectric domain walls (DWs) acting as a series of naturally occurring $p$-$n$ junctions \cite{seidel:2011:efficient}.
However, later experiments \cite{bhatnagar:2013:role} and first-principles calculations of the bulk photovoltaic effect (BPVE) in BiFeO$_3$ \cite{young:2012:firstprinciples}.
show that the observed large photovoltage of BiFeO$_3$ can be explained on the basis of the BPVE alone, with no need for DW contributions.
This means that the role and function of DWs with respect to the photovoltage is still not clear.
In Refs.~[\onlinecite{seidel:2011:efficient}] and [\onlinecite{yang:2010:above}] a model for the DW contribution to the photovoltage was proposed (see Fig~\ref{fig:model_seidel}),
where discontinuities in the ferroelectric polarization at the DWs lead to electrostatic potential steps.
These separate the photogenerated charges and trap electrons and holes on opposite sides of the DWs,
thus impeding charge-carrier recombination by spatial separation.
First-principles calculations using density-functional theory (DFT) were performed 
to investigate neutral ferroelectric DWs in BiFeO$_3$ \cite{seidel:2009:conduction,lubk:2009:first,dieguez:2013:domain,wang:2013:bifeo3,ren:2013:ferroelectric,chen:2017:polar},
and the electronic potential at the walls was investigated 
using indirect methods \cite{seidel:2009:conduction,lubk:2009:first} and without explicitly considering excess charge carriers.
Such approach
 was first adopted with first-principles calculations for
 neutral 90\textdegree~\new{DW} in PbTiO$_3$ \cite{meyer:2002:ab} and yielded electrostatic potential steps of about 0.18~eV at the walls.
  The same approach applied to BiFeO$_3$ \cite{seidel:2009:conduction} yielded potential steps of 0.02--0.18~eV depending on the wall type.
  In contrast, Ginzburg-Landau-Devonshire theory \cite{marton:2010:domain} combined with flexoelectric coupling and deformation-potential theory 
  for BiFeO$_3$ \cite{morozovska:2012:anisotropic} 
  yielded bell-shaped, symmetric electrostatic potentials and charge-carrier distributions at the walls.
Here we show that it is necessary to go beyond these approaches and to both calculate directly the electronic potential and take explicitly into account
excess charges to reveal the unusual charge localization behavior and the nature of the electronic zigzag potential
in BiFeO$_3$.
We find important deviations in the charge-density distribution and the electrostatic potential with respect to the previously proposed model:
excess electrons are indeed trapped at the \new{DW}, but holes are strongly delocalized,
leading to an asymmetric charge density distribution.
As a consequence the potential has indeed a zigzag profile, but this largely originates from the trapped excess electrons.
The previously proposed model is depicted in Fig.~\ref{fig:model_seidel}, and our suggested modification is shown in Fig.~\ref{fig:model}.
%
%
%
\paragraph*{\label{sec:methods}Methods.}
In rhombohedral perovskites, such as BiFeO$_3$, the polarization in adjacent domains can form angles of about 71\textdegree, 109\textdegree, and 180\textdegree,
which are all studied here. For each DW angle, we selected the DW with the lowest possible Miller indices which is electrically neutral and mechanically compatible \cite{fousek:1969:orientation,dieguez:2013:domain}.
The DFT calculations were performed with the {\sc{vasp}} code \cite{kresse:1996:efficiency},
using the projector-augmented wave method and pseudopotentials with 5 (Bi), 16 (Fe), and 6 (O) valence electrons, respectively.
We used the local-density approximation combined with a Hubbard-$U$ term of 5.3~eV following Dudarev's
scheme \cite{dudarev:1998:electron}.
This $U$ value was taken from the materials project \cite{jain:2013:materials} and is optimized for oxide formation energies,
but also yields band gaps close to experiments.
The reciprocal space was sampled with $2\times5\times3$ $k$ points for the 71\textdegree\ and 180\textdegree\ DW,
and with $2\times5\times5$ $k$ points for the 109\textdegree\ wall.
Plane-wave basis functions with energies up to 520~eV were used.
We employed a supercell approach with periodic boundary conditions,
such that each supercell contains 120 atoms and two \new{DW}.
Both the atomic positions and cell parameters were allowed to relax
until the energy difference between ionic relaxation steps fell below 0.1~meV.
We use a coordinate system with axes \{$\v{e}_r$, $\v{e}_s$, $\v{e}_t$\},
where $\v{e}_r\parallel P_r$ is the polarization component which changes sign at the \new{DW},
$\v{e}_s$ is perpendicular to the \new{DW} plane, and $\v{e}_t=\v{e}_r\times\v{e}_s$.
The polarization profiles were calculated from the ionic positions $\v{u}_i$ and the formal ionic charges, $Z_i$ (Bi$^{3+}$, Fe$^{3+}$ and O$^{2-}$),
weighted by $w_i$ ($w_{\rm Bi}=1/8$, $w_{\rm Fe}=1$, and $w_{\rm O}=1/2$)
for each Fe-centered five-atom perovskite cell as
\begin{equation}
  \label{eq:pol}
  \v{P}=\sum_i w_i Z_i\v{u}_i\:.
\end{equation}
In order to investigate the localization of the excess charges, we calculated their densities as
\begin{equation}\label{eq:rho}
  \varrho_{\textrm{excess}}=\left\{ {\sum_{c\v{k}} |\psi_{c\v{k}}|^2\,f_{c\v{k}}\textrm{\ (electrons)}}\atop{\sum_{v\v{k}} |\psi_{v\v{k}}|^2\,(1-f_{v\v{k}})\textrm{\ (holes),}} \right.
\end{equation}
where $f_{v\v{k}}$ is the occupation number of the Bloch function $\psi_{v\v{k}}$, and $v$ and $c$ are
the valence- and conduction-band indices, respectively.
\new{
 Symmetry causes the trap states to be only half occupied, if only one excess electron is introduced in the 
supercell. In order to avoid such artificial partial occupation, we placed two electrons or holes in the supercell.
In test calculations without symmetry and with a single excess electron only,
the electron localized at one of the two \new{DW}, {\it i.e.} the trap-state occupation was the same as that
of the supercell with two excess electrons and symmetry.
}
%
\paragraph*{Results and discussion.}
\begin{figure}
   \def\svgwidth{0.31\textwidth}
   \input{DW_structure_71.pdf_tex}
   \includegraphics[width=0.5\textwidth]{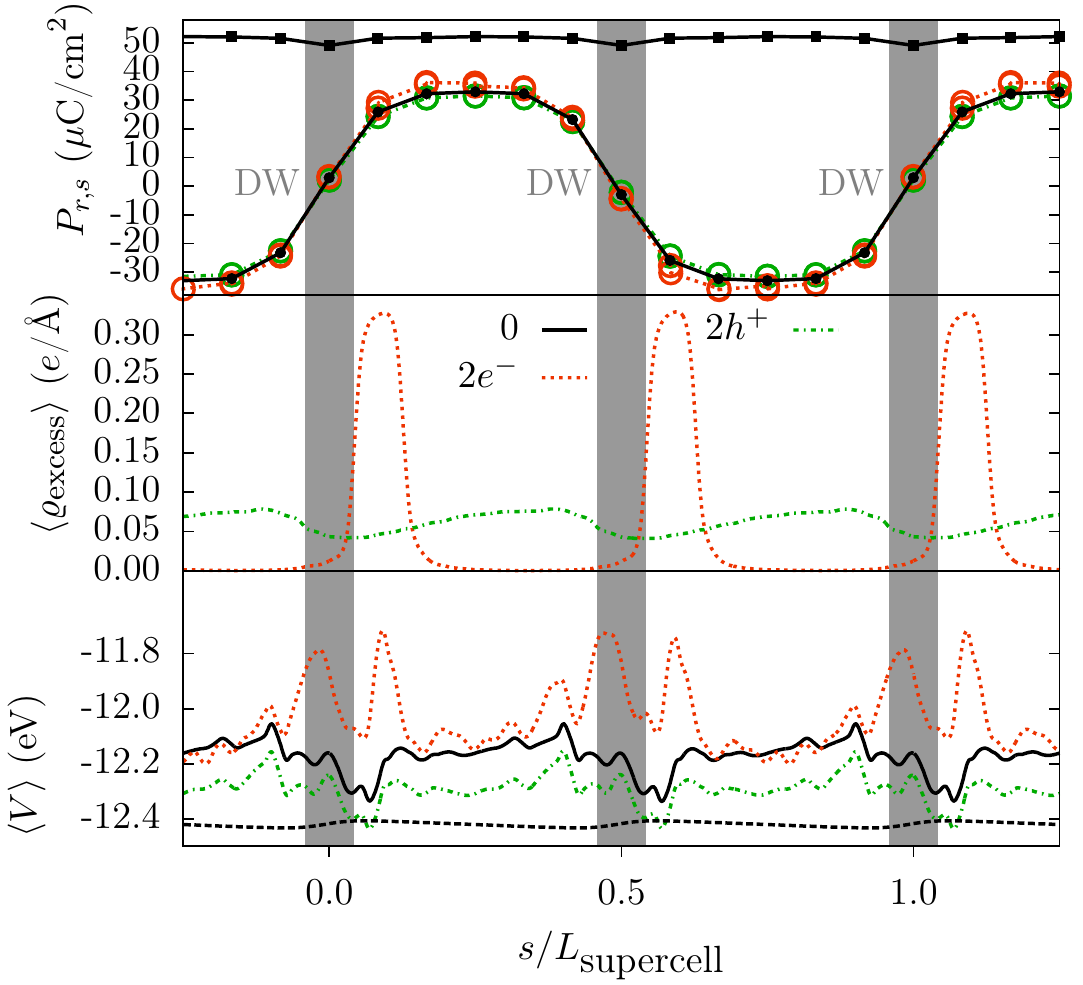}
   \caption{\label{fig:pol_charge_pot_71} (Color online) Characterization of the 71\textdegree~DW.
     Ferroelectric polarization profile $P_r$ (top), charge density $\varrho_{\textrm{excess}}$ of excess 
     electrons and holes (center), and electronic potential $V$ (bottom). $P_r$ is the $r$ component of 
     the polarization. $\varrho_{\textrm{excess}}$ and $V$ are averaged over one atomic layer to smoothen 
     strong rapid oscillations at atomic nuclei \cite{meyer:2002:ab}. Results are presented for the neutral cell 
     (solid black line), and for those containing two extra electrons, $2e^-$ (dashed orange line), or two extra holes, 
     $2h^+$ (dash-dotted green line). The dashed black line in the bottom panel represents the electronic 
     potential according to Eq.~(\ref{eq:pot}).
  }
\end{figure}
\begin{figure}[t]
   \def\svgwidth{0.30\textwidth}
   \input{DW_structure_109.pdf_tex}
   \includegraphics[width=0.5\textwidth]{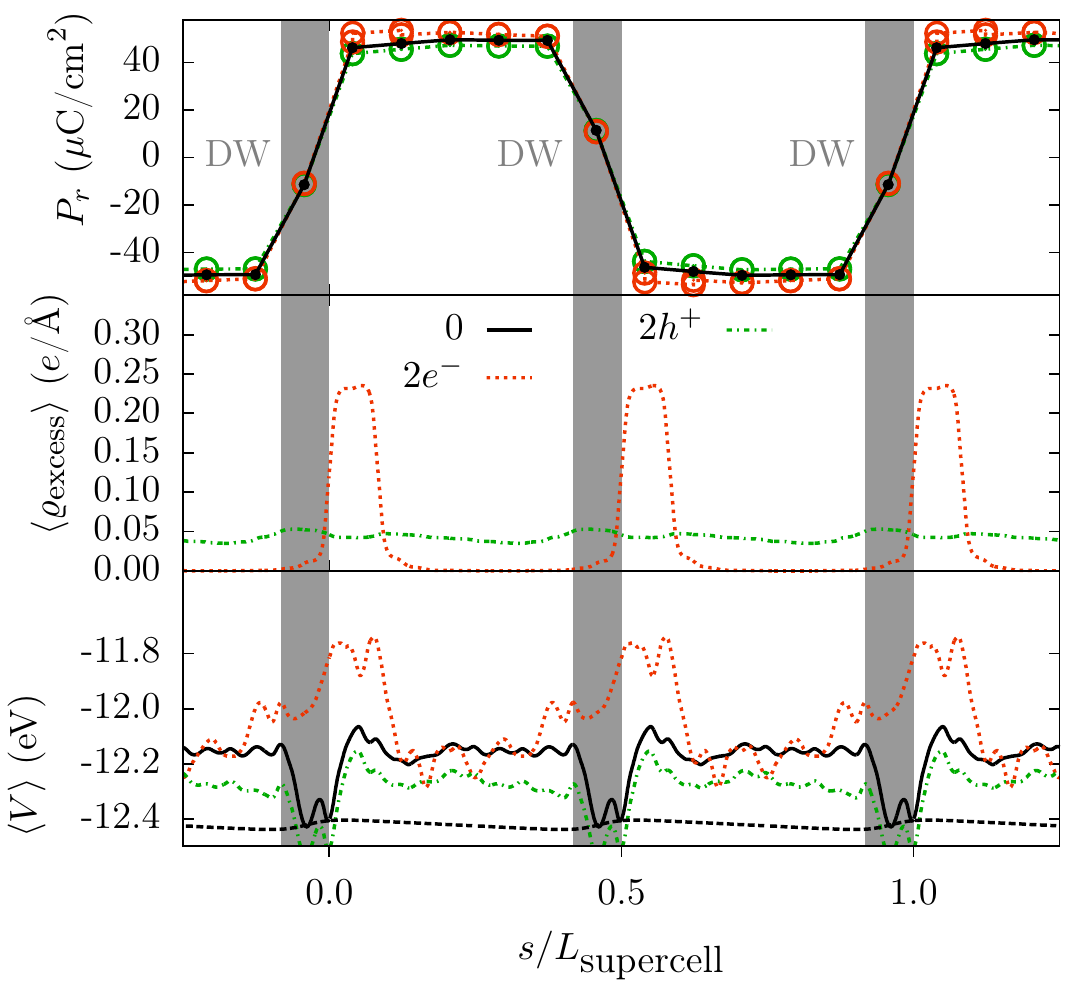}
   \caption{\label{fig:pol_charge_pot_109} (Color online) The same as Fig.~\ref{fig:pol_charge_pot_71} for the 109\textdegree~DW. 
  }
\end{figure}
%
In the top panels of Figs.~\ref{fig:pol_charge_pot_71}--\ref{fig:pol_charge_pot_180} we show the polarization profile, $P_r$, across the DW, 
calculated from Eq.~(\ref{eq:pol}).
Clearly for all DWs the polarization profile remains unchanged regardless of the system's total charge (neutral, positively and negatively charged), meaning that the polar state of the DW structure
is insensitive to charging.
\begin{figure}[t]
   \def\svgwidth{0.315\textwidth}
   \input{DW_structure_180.pdf_tex}
   \includegraphics[width=0.5\textwidth]{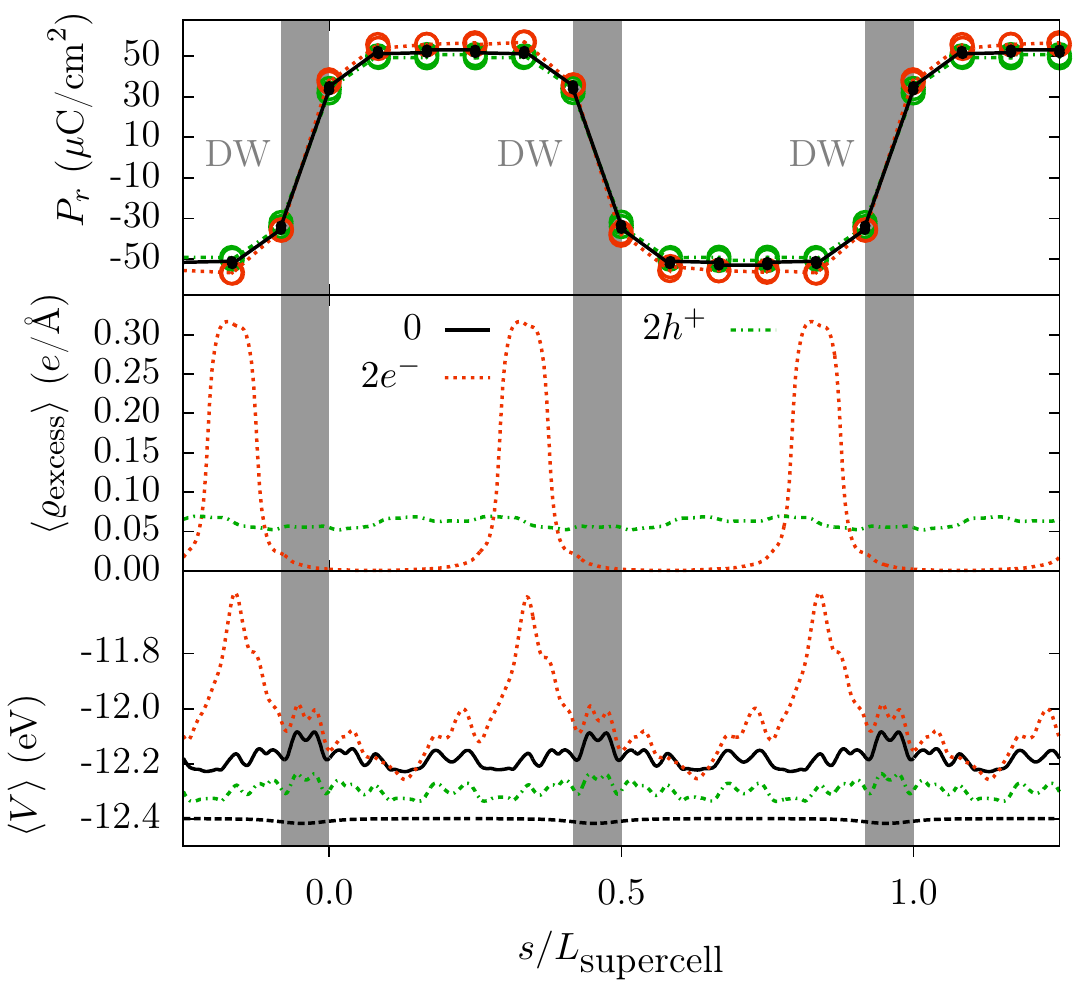}
   \caption{\label{fig:pol_charge_pot_180} (Color online) The same as Figs.~\ref{fig:pol_charge_pot_71} and \ref{fig:pol_charge_pot_109} for the 180\textdegree~DW. 
  }
\end{figure}
In the center panels we show the excess charge density profile
upon injection of holes or electrons.
In the case of electrons such additional charge localizes tightly at the DWs. 
In contrast the positive hole charge \new{is distributed} over the entire cell. For the 71\textdegree~DW one
can still observe some moderate hole localization at the DW side opposite to that where the electrons accumulate, but this is
not visible for the 109\textdegree~and 180\textdegree~DWs.
The potential resulting from such charge distribution is presented next in the bottom panels (here we show the full Kohn-Sham potential).
For comparison, we also display the classical electrostatic potential estimated as in 
Ref.~[\onlinecite{meyer:2002:ab}] from the tiny \new{reduction in} the normal component of the 
polarization profile  $\Delta P_s$:
\begin{equation}
  \label{eq:pot}
  V_{\rm es}=\frac{-e}{\varepsilon\varepsilon_0}\int_{s^-}^{s^+}\Delta P_s(s)\textrm{d}s,
\end{equation}
\noindent where  $\epsilon \approx 25$ is our calculated bulk electronic permittivity, $\varepsilon_0$ is the 
vacuum permittivity, and $s^-$ and $s^+$ are positions left and right of the wall. However, this is too crude 
an approximation to capture any systematic trends in the real profiles of the full Kohn-Sham potential in the 
present case. Let us focus first on the Kohn-Sham potential of the neutral system (solid black line).
For the 71\textdegree~and 109\textdegree~DWs the dominant feature in the potential is a narrow local 
minimum, which extends over up to two perovskite monolayers at the DW. 
In the case of the 180\textdegree\ DW the potential well becomes a shallow barrier.
Similar potential wells/barriers have been found as a result of flexoelectric coupling and/or deformation potential in Ref.~\onlinecite{morozovska:2012:anisotropic}.
Despite the differences in the potential
among the three DWs, excess charges distribute similarly, as described above. The charge accumulation in the case of excess electrons
modifies the potential, which is now strongly repulsive at the DW (see dashed orange lines). In contrast, hole doping, where the positive charge
distributes across the whole cell, affects little the general shape of the potential and its main effect is an approximately rigid energy shift
(dash-dotted green lines). Thus we conclude that in a situation of excess of electrons, which can arise from intentional doping, intrinsic defects, or photo-carrier generation, the potential will present a zigzag profile along the cell.
Our calculation has shown that it is possible to store about one electron per DW in our unit cell, which has a high DW density. Assuming the same excess-electron density per DW area, $A_{\rm DW}$,
and a more realistic \new{DW} separation $d_{\rm DW}$ of 140~nm \cite{seidel:2011:efficient}, we arrive 
at a corresponding excess-electron density of 
$n_e=1/(A_{\rm DW}   d_{\rm DW}) \approx 1.7\cdot 10^{19}{\rm\ cm}^{-3}$, which could be entirely stored 
in the walls.

\begin{figure}
  \includegraphics[width=0.5\textwidth]{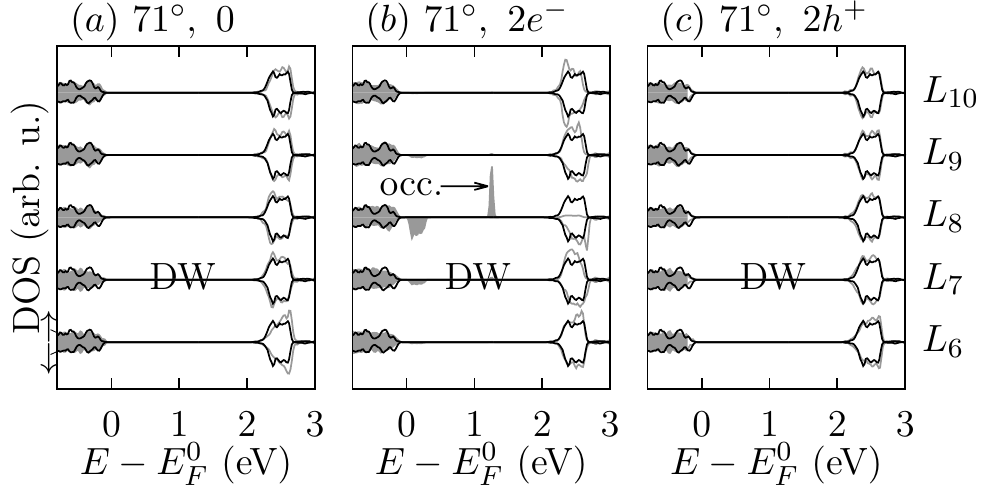}
  \includegraphics[width=0.5\textwidth]{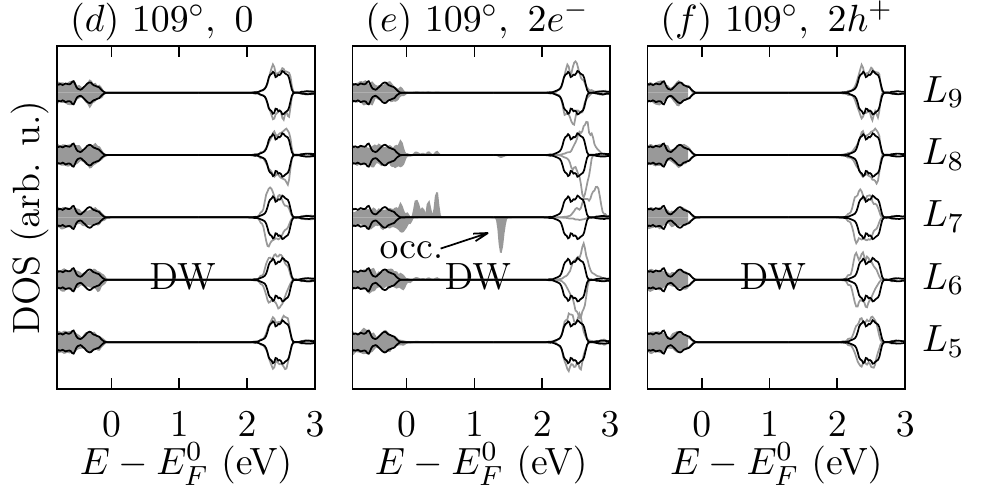}
  \includegraphics[width=0.5\textwidth]{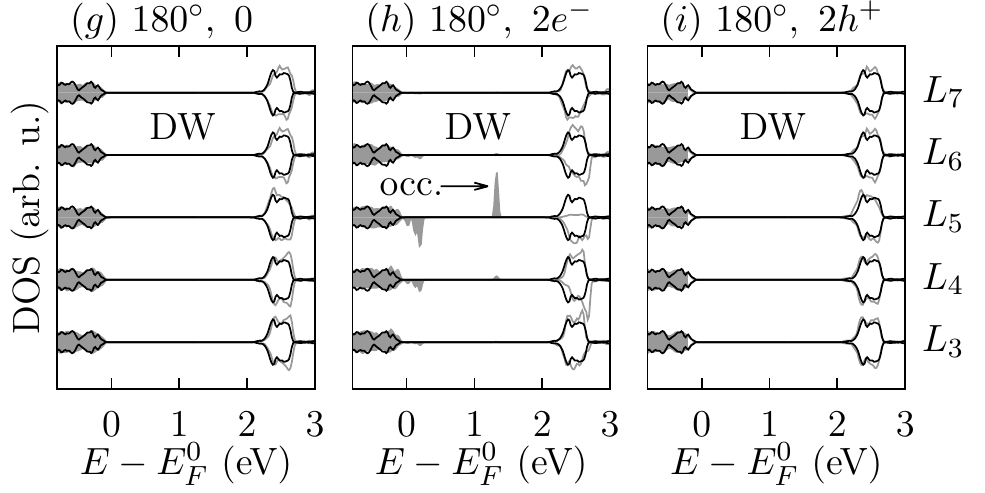}
  \caption{\label{fig:dos}Density of states (DOS) of the \new{DW} system (gray) and the charge-neutral monodomain system (black) projected 
    on the atomic layers. (a), (d), (g) Charge-neutral \new{DW}; (b), (e), (h) DW with electrons; (c), (f), (i) DW with holes. 
    (a)--(c) 71\textdegree\ \new{DW};  (d)--(f) 109\textdegree\ DW; (g)--(i) 180\textdegree\ \new{DW}.
  $E^0_F$ is the Fermi energy 
of the charge-neutral system. 
Occupied levels are shaded in gray. 
\new{DOS} for positive spin ($\uparrow$) and negative spin ($\downarrow$) are both shown,
  the $\downarrow$ DOS was multiplied by --1. DOS were vertically shifted for better visibility.
}
\end{figure}

The effect of the DWs and of the charge accumulation on the electronic structure is analyzed next in Fig.~\ref{fig:dos},
where we show the layer-resolved density of states (DOS) for the different charging situations investigated here.
For both the neutral and the positively charged systems the DOS appears little perturbed by the presence of the DW, and it is essentially identical
regardless of the layer on which it is projected. In contrast, when excess electrons are introduced, one can clearly observe the formation of two
narrow peaks in the DOS localized at the layers adjacent to the DW.
The gap level close to the valence band originates from occupied Fe $d$ orbitals oriented along Fe-O bonds being pushed energetically
upward, while the one at higher energy is from the Fe $d$ orbitals that are directed away from Fe-O bonds and are occupied by the excess electron.
The same electron trapping (small electron polaron) has been discussed in the case of charged domain 
walls in ErMnO$_3$ \cite{mundy:2017:functional} and PbTiO$_3$ \cite{rahmanizadeh:2014:charge}.
In the case of BiFeO$_3$, we could have actually expected electron trapping instead of delocalized metallic 
behavior because the $d$ states of Fe at the bottom of the conduction band are relatively isolated both in 
space and in energy. 
%
Let us stress that the electronic polaron trap state, present here, should be distinguished from the 
bending (down-shift) of the entire conduction band, the latter being encountered in the case of the 
head-to-head, {\it i.e.} nominally charged \new{DW} \cite{sluka:2013:free,bednyakov:2015:formation,liu:2015:ferroelectric,liu:2017:stable,li:2018:resonant}.
\new{In the case of charged \new{DW}, the large change in $P_s$ (normal to the wall)
 should result in a large bound charge and a strong electrostatic potential 
variation at the wall, such that the electrostatic potential would be 
the dominant factor governing electron localization at the wall, 
and other factors like strain or tilt variations would play a minor role.  
In contrast, in the case of the neutral DW, electrostatic potential variations, if present,  
are relatively small, and other effects like strain or tilt variations can become more important.
}
\new{The reason why the excess electrons localize on one side of the wall appears to 
be related to strain and/or tilt effects. We notice in the case of the 71\textdegree~ and the 109\textdegree~wall
that before adding the excess electron, the octahedral tilt is already slightly larger on the side of the wall where the excess electron 
finally localizes. 
Once the electron polaron forms, the octahedral cage surrounding it expands, and the 
tilt increases further, thus lowering the energy of the trapped electron.
In the case of the 180\textdegree~wall, the electron also localizes on a site with large tilt,  
but here the left and right side of the wall are equivalent, and the electron might as well have localized on the other side.}
\new{The} electronic trap state might be 
detectable in photoemission experiments.

In both Refs.~[\onlinecite{bhatnagar:2013:role}] and [\onlinecite{yang:2010:above}], the \new{DW} in BiFeO$_3$ were found to be more conductive than
the domain interior. Such \new{DW} conductivity appears to be thermally activated \cite{bhatnagar:2013:role}, and trap states about 1~eV below the conduction
band are involved in the photocurrent generation \cite{yang:2015:electronic}. The deep levels which we find at electron-doped DWs could provide such
trap states.

In this study we consider the idealized case of pristine DWs,
which is well defined, but incomplete. Real samples contain
various amounts of oxygen vacancies and/or other point defects, which
may aggregate at the DWs. These may modify the electronic potential and the amount of excess
electrons by acting as traps. Therefore a fully realistic picture will be obtained by considering pristine \new{DW} sections, intersected by point defects.
Notably, the presence of point defects at \new{DW} could also influence the resistivity of the walls themselves.
The trap states at the \new{DW} ($\approx 1$~eV below the conduction band) are apparently deeper than those in the domain interior 
  ($\approx 0.6$~eV below the conduction band \cite{clark:2009:energy}), but
  possibly shallower than oxygen vacancy states at the wall. Therefore oxygen vacancies at the wall may compete with the \new{DW} states as electron traps.
%
\paragraph*{Summary and conclusion.}
%
\new{Neutral} ferroelectric domain walls in BiFeO$_3$ strongly trap excess electrons, but only weakly attract holes.
The different localization behavior of electrons and holes at the walls may be understood
on the basis of the electronic states which dominate the top of the valence band (delocalized O $p$ states)
and the bottom of the conduction band (localized Fe $d$ states).
\par
The potential profile at the domain walls is dominated by excess charge carriers.
Without excess electrons, the electronic potential at the domain wall exhibits mainly a well or barrier.
Once electrons are trapped in the domain walls, 
\new{they create a strong repulsive zigzag potential,} 
whose amplitude depends on
excess-electron and domain-wall density.
\new{The potential at} the 71\textdegree~and 109\textdegree~domain walls is asymmetric,
which could enable a net photovoltaic current generation, 
\new{whereas the potential} at the 180\textdegree~domain
wall is approximately symmetric, indicating that this wall may be photovoltaically less active.
%
The trap states for electrons at domain walls in BiFeO$_3$ may be at the origin of the thermally activated domain-wall conduction found
in experiments.
\\
\paragraph*{Acknowledgments.}
This project  has  received  funding  from  the
European Union's Horizon 2020 research and innovation programme under the Marie Sk\l{}odowska-Curie
Grant Agreement No. 746964--FERROVOLT and from the
Czech Science Foundation (Project No. 15-04121S).
Computation time was provided by the Czech Metacentrum
and the Trinity Centre for High Performance Computing funded by Science Foundation Ireland.
Figures were made using {\sc{gnuplot}}, {\sc{inkscape}}, and {\sc VESTA} \cite{momma:2011:vesta}.
Pavel M\'{a}rton brought this collaboration together.

\bibliography{jabbr,all}

\begin{thebibliography}{27}%
\makeatletter
\providecommand \@ifxundefined [1]{%
 \@ifx{#1\undefined}
}%
\providecommand \@ifnum [1]{%
 \ifnum #1\expandafter \@firstoftwo
 \else \expandafter \@secondoftwo
 \fi
}%
\providecommand \@ifx [1]{%
 \ifx #1\expandafter \@firstoftwo
 \else \expandafter \@secondoftwo
 \fi
}%
\providecommand \natexlab [1]{#1}%
\providecommand \enquote  [1]{``#1''}%
\providecommand \bibnamefont  [1]{#1}%
\providecommand \bibfnamefont [1]{#1}%
\providecommand \citenamefont [1]{#1}%
\providecommand \href@noop [0]{\@secondoftwo}%
\providecommand \href [0]{\begingroup \@sanitize@url \@href}%
\providecommand \@href[1]{\@@startlink{#1}\@@href}%
\providecommand \@@href[1]{\endgroup#1\@@endlink}%
\providecommand \@sanitize@url [0]{\catcode `\\12\catcode `\$12\catcode
  `\&12\catcode `\#12\catcode `\^12\catcode `\_12\catcode `\%12\relax}%
\providecommand \@@startlink[1]{}%
\providecommand \@@endlink[0]{}%
\providecommand \url  [0]{\begingroup\@sanitize@url \@url }%
\providecommand \@url [1]{\endgroup\@href {#1}{\urlprefix }}%
\providecommand \urlprefix  [0]{URL }%
\providecommand \Eprint [0]{\href }%
\providecommand \doibase [0]{http://dx.doi.org/}%
\providecommand \selectlanguage [0]{\@gobble}%
\providecommand \bibinfo  [0]{\@secondoftwo}%
\providecommand \bibfield  [0]{\@secondoftwo}%
\providecommand \translation [1]{[#1]}%
\providecommand \BibitemOpen [0]{}%
\providecommand \bibitemStop [0]{}%
\providecommand \bibitemNoStop [0]{.\EOS\space}%
\providecommand \EOS [0]{\spacefactor3000\relax}%
\providecommand \BibitemShut  [1]{\csname bibitem#1\endcsname}%
\let\auto@bib@innerbib\@empty
\bibitem [{\citenamefont {Seidel}\ \emph {et~al.}(2011)\citenamefont {Seidel},
  \citenamefont {Fu}, \citenamefont {Yang}, \citenamefont
  {Alarc{\'o}n-Llad{\'o}}, \citenamefont {Wu}, \citenamefont {Ramesh},\ and\
  \citenamefont {Ager~III}}]{seidel:2011:efficient}%
  \BibitemOpen
  \bibfield  {author} {\bibinfo {author} {\bibfnamefont {J.}~\bibnamefont
  {Seidel}}, \bibinfo {author} {\bibfnamefont {D.}~\bibnamefont {Fu}}, \bibinfo
  {author} {\bibfnamefont {S.-Y.}\ \bibnamefont {Yang}}, \bibinfo {author}
  {\bibfnamefont {E.}~\bibnamefont {Alarc{\'o}n-Llad{\'o}}}, \bibinfo {author}
  {\bibfnamefont {J.}~\bibnamefont {Wu}}, \bibinfo {author} {\bibfnamefont
  {R.}~\bibnamefont {Ramesh}}, \ and\ \bibinfo {author} {\bibfnamefont {J.~W.}\
  \bibnamefont {Ager~III}},\ }\href@noop {} {\bibfield  {journal} {\bibinfo
  {journal} {Physical review letters}\ }\textbf {\bibinfo {volume} {107}},\
  \bibinfo {pages} {126805} (\bibinfo {year} {2011})}\BibitemShut {NoStop}%
\bibitem [{\citenamefont {Bhatnagar}\ \emph {et~al.}(2013)\citenamefont
  {Bhatnagar}, \citenamefont {Chaudhuri}, \citenamefont {Kim}, \citenamefont
  {Hesse},\ and\ \citenamefont {Alexe}}]{bhatnagar:2013:role}%
  \BibitemOpen
  \bibfield  {author} {\bibinfo {author} {\bibfnamefont {A.}~\bibnamefont
  {Bhatnagar}}, \bibinfo {author} {\bibfnamefont {A.~R.}\ \bibnamefont
  {Chaudhuri}}, \bibinfo {author} {\bibfnamefont {Y.~H.}\ \bibnamefont {Kim}},
  \bibinfo {author} {\bibfnamefont {D.}~\bibnamefont {Hesse}}, \ and\ \bibinfo
  {author} {\bibfnamefont {M.}~\bibnamefont {Alexe}},\ }\href@noop {}
  {\bibfield  {journal} {\bibinfo  {journal} {{Nature Communications}}\
  }\textbf {\bibinfo {volume} {4}},\ \bibinfo {pages} {2835} (\bibinfo {year}
  {2013})}\BibitemShut {NoStop}%
\bibitem [{\citenamefont {Young}\ \emph {et~al.}(2012)\citenamefont {Young},
  \citenamefont {Zheng},\ and\ \citenamefont
  {Rappe}}]{young:2012:firstprinciples}%
  \BibitemOpen
  \bibfield  {author} {\bibinfo {author} {\bibfnamefont {S.~M.}\ \bibnamefont
  {Young}}, \bibinfo {author} {\bibfnamefont {F.}~\bibnamefont {Zheng}}, \ and\
  \bibinfo {author} {\bibfnamefont {A.~M.}\ \bibnamefont {Rappe}},\ }\href@noop
  {} {\bibfield  {journal} {\bibinfo  {journal} {Phys. Rev. Lett.}\ }\textbf
  {\bibinfo {volume} {109}},\ \bibinfo {pages} {236601} (\bibinfo {year}
  {2012})}\BibitemShut {NoStop}%
\bibitem [{\citenamefont {Yang}\ \emph {et~al.}(2010)\citenamefont {Yang},
  \citenamefont {Seidel}, \citenamefont {Byrnes}, \citenamefont {Shafer},
  \citenamefont {Yang}, \citenamefont {Rossell}, \citenamefont {Yu},
  \citenamefont {Chu}, \citenamefont {Scott}, \citenamefont {III},
  \citenamefont {Martin},\ and\ \citenamefont {Ramesh}}]{yang:2010:above}%
  \BibitemOpen
  \bibfield  {author} {\bibinfo {author} {\bibfnamefont {S.~Y.}\ \bibnamefont
  {Yang}}, \bibinfo {author} {\bibfnamefont {J.}~\bibnamefont {Seidel}},
  \bibinfo {author} {\bibfnamefont {S.~J.}\ \bibnamefont {Byrnes}}, \bibinfo
  {author} {\bibfnamefont {P.}~\bibnamefont {Shafer}}, \bibinfo {author}
  {\bibfnamefont {C.-H.}\ \bibnamefont {Yang}}, \bibinfo {author}
  {\bibfnamefont {M.~D.}\ \bibnamefont {Rossell}}, \bibinfo {author}
  {\bibfnamefont {P.}~\bibnamefont {Yu}}, \bibinfo {author} {\bibfnamefont
  {Y.-H.}\ \bibnamefont {Chu}}, \bibinfo {author} {\bibfnamefont {J.~F.}\
  \bibnamefont {Scott}}, \bibinfo {author} {\bibfnamefont {J.~W.~A.}\
  \bibnamefont {III}}, \bibinfo {author} {\bibfnamefont {L.~W.}\ \bibnamefont
  {Martin}}, \ and\ \bibinfo {author} {\bibfnamefont {R.}~\bibnamefont
  {Ramesh}},\ }\href@noop {} {\bibfield  {journal} {\bibinfo  {journal} {Nature
  Nanotechnology}\ }\textbf {\bibinfo {volume} {5}},\ \bibinfo {pages} {143}
  (\bibinfo {year} {2010})}\BibitemShut {NoStop}%
\bibitem [{\citenamefont {Seidel}\ \emph {et~al.}(2009)\citenamefont {Seidel},
  \citenamefont {Martin}, \citenamefont {He}, \citenamefont {Zhan},
  \citenamefont {Chu}, \citenamefont {Rother}, \citenamefont {Hawkridge},
  \citenamefont {Maksymovych}, \citenamefont {Yu}, \citenamefont {Gajek} \emph
  {et~al.}}]{seidel:2009:conduction}%
  \BibitemOpen
  \bibfield  {author} {\bibinfo {author} {\bibfnamefont {J.}~\bibnamefont
  {Seidel}}, \bibinfo {author} {\bibfnamefont {L.~W.}\ \bibnamefont {Martin}},
  \bibinfo {author} {\bibfnamefont {Q.}~\bibnamefont {He}}, \bibinfo {author}
  {\bibfnamefont {Q.}~\bibnamefont {Zhan}}, \bibinfo {author} {\bibfnamefont
  {Y.-H.}\ \bibnamefont {Chu}}, \bibinfo {author} {\bibfnamefont
  {A.}~\bibnamefont {Rother}}, \bibinfo {author} {\bibfnamefont
  {M.}~\bibnamefont {Hawkridge}}, \bibinfo {author} {\bibfnamefont
  {P.}~\bibnamefont {Maksymovych}}, \bibinfo {author} {\bibfnamefont
  {P.}~\bibnamefont {Yu}}, \bibinfo {author} {\bibfnamefont {M.}~\bibnamefont
  {Gajek}},  \emph {et~al.},\ }\href@noop {} {\bibfield  {journal} {\bibinfo
  {journal} {Nat. Mater.}\ }\textbf {\bibinfo {volume} {8}},\ \bibinfo {pages}
  {229} (\bibinfo {year} {2009})}\BibitemShut {NoStop}%
\bibitem [{\citenamefont {Lubk}\ \emph {et~al.}(2009)\citenamefont {Lubk},
  \citenamefont {Gemming},\ and\ \citenamefont {Spaldin}}]{lubk:2009:first}%
  \BibitemOpen
  \bibfield  {author} {\bibinfo {author} {\bibfnamefont {A.}~\bibnamefont
  {Lubk}}, \bibinfo {author} {\bibfnamefont {S.}~\bibnamefont {Gemming}}, \
  and\ \bibinfo {author} {\bibfnamefont {N.~A.}\ \bibnamefont {Spaldin}},\
  }\href@noop {} {\bibfield  {journal} {\bibinfo  {journal} {Phys. Rev. B}\
  }\textbf {\bibinfo {volume} {80}},\ \bibinfo {pages} {104110} (\bibinfo
  {year} {2009})}\BibitemShut {NoStop}%
\bibitem [{\citenamefont {Di{\'e}guez}\ \emph {et~al.}(2013)\citenamefont
  {Di{\'e}guez}, \citenamefont {Aguado-Puente}, \citenamefont {Junquera},\ and\
  \citenamefont {{\'I}{\~n}iguez}}]{dieguez:2013:domain}%
  \BibitemOpen
  \bibfield  {author} {\bibinfo {author} {\bibfnamefont {O.}~\bibnamefont
  {Di{\'e}guez}}, \bibinfo {author} {\bibfnamefont {P.}~\bibnamefont
  {Aguado-Puente}}, \bibinfo {author} {\bibfnamefont {J.}~\bibnamefont
  {Junquera}}, \ and\ \bibinfo {author} {\bibfnamefont {J.}~\bibnamefont
  {{\'I}{\~n}iguez}},\ }\href@noop {} {\bibfield  {journal} {\bibinfo
  {journal} {Phys. Rev. B}\ }\textbf {\bibinfo {volume} {87}},\ \bibinfo
  {pages} {024102} (\bibinfo {year} {2013})}\BibitemShut {NoStop}%
\bibitem [{\citenamefont {Wang}\ \emph {et~al.}(2013)\citenamefont {Wang},
  \citenamefont {Nelson}, \citenamefont {Melville}, \citenamefont {Winchester},
  \citenamefont {Shang}, \citenamefont {Liu}, \citenamefont {Schlom},
  \citenamefont {Pan},\ and\ \citenamefont {Chen}}]{wang:2013:bifeo3}%
  \BibitemOpen
  \bibfield  {author} {\bibinfo {author} {\bibfnamefont {Y.}~\bibnamefont
  {Wang}}, \bibinfo {author} {\bibfnamefont {C.}~\bibnamefont {Nelson}},
  \bibinfo {author} {\bibfnamefont {A.}~\bibnamefont {Melville}}, \bibinfo
  {author} {\bibfnamefont {B.}~\bibnamefont {Winchester}}, \bibinfo {author}
  {\bibfnamefont {S.}~\bibnamefont {Shang}}, \bibinfo {author} {\bibfnamefont
  {Z.-K.}\ \bibnamefont {Liu}}, \bibinfo {author} {\bibfnamefont {D.~G.}\
  \bibnamefont {Schlom}}, \bibinfo {author} {\bibfnamefont {X.}~\bibnamefont
  {Pan}}, \ and\ \bibinfo {author} {\bibfnamefont {L.-Q.}\ \bibnamefont
  {Chen}},\ }\href@noop {} {\bibfield  {journal} {\bibinfo  {journal} {Phys.
  Rev. Lett.}\ }\textbf {\bibinfo {volume} {110}},\ \bibinfo {pages} {267601}
  (\bibinfo {year} {2013})}\BibitemShut {NoStop}%
\bibitem [{\citenamefont {Ren}\ \emph {et~al.}(2013)\citenamefont {Ren},
  \citenamefont {Yang}, \citenamefont {Di\'eguez}, \citenamefont {\'I\~niguez},
  \citenamefont {Choudhury},\ and\ \citenamefont
  {Bellaiche}}]{ren:2013:ferroelectric}%
  \BibitemOpen
  \bibfield  {author} {\bibinfo {author} {\bibfnamefont {W.}~\bibnamefont
  {Ren}}, \bibinfo {author} {\bibfnamefont {Y.}~\bibnamefont {Yang}}, \bibinfo
  {author} {\bibfnamefont {O.}~\bibnamefont {Di\'eguez}}, \bibinfo {author}
  {\bibfnamefont {J.}~\bibnamefont {\'I\~niguez}}, \bibinfo {author}
  {\bibfnamefont {N.}~\bibnamefont {Choudhury}}, \ and\ \bibinfo {author}
  {\bibfnamefont {L.}~\bibnamefont {Bellaiche}},\ }\href {\doibase
  10.1103/PhysRevLett.110.187601} {\bibfield  {journal} {\bibinfo  {journal}
  {Phys. Rev. Lett.}\ }\textbf {\bibinfo {volume} {110}},\ \bibinfo {pages}
  {187601} (\bibinfo {year} {2013})}\BibitemShut {NoStop}%
\bibitem [{\citenamefont {Chen}\ \emph {et~al.}(2017)\citenamefont {Chen},
  \citenamefont {Kuo},\ and\ \citenamefont {Chew}}]{chen:2017:polar}%
  \BibitemOpen
  \bibfield  {author} {\bibinfo {author} {\bibfnamefont {Y.-W.}\ \bibnamefont
  {Chen}}, \bibinfo {author} {\bibfnamefont {J.-L.}\ \bibnamefont {Kuo}}, \
  and\ \bibinfo {author} {\bibfnamefont {K.-H.}\ \bibnamefont {Chew}},\
  }\href@noop {} {\bibfield  {journal} {\bibinfo  {journal} {J. Appl. Phys.}\
  }\textbf {\bibinfo {volume} {122}},\ \bibinfo {pages} {075103} (\bibinfo
  {year} {2017})}\BibitemShut {NoStop}%
\bibitem [{\citenamefont {Meyer}\ and\ \citenamefont
  {Vanderbilt}(2002)}]{meyer:2002:ab}%
  \BibitemOpen
  \bibfield  {author} {\bibinfo {author} {\bibfnamefont {B.}~\bibnamefont
  {Meyer}}\ and\ \bibinfo {author} {\bibfnamefont {D.}~\bibnamefont
  {Vanderbilt}},\ }\href@noop {} {\bibfield  {journal} {\bibinfo  {journal}
  {Phys. Rev. B}\ }\textbf {\bibinfo {volume} {65}},\ \bibinfo {pages} {104111}
  (\bibinfo {year} {2002})}\BibitemShut {NoStop}%
\bibitem [{\citenamefont {Marton}\ \emph {et~al.}(2010)\citenamefont {Marton},
  \citenamefont {Rychetsky},\ and\ \citenamefont
  {Hlinka}}]{marton:2010:domain}%
  \BibitemOpen
  \bibfield  {author} {\bibinfo {author} {\bibfnamefont {P.}~\bibnamefont
  {Marton}}, \bibinfo {author} {\bibfnamefont {I.}~\bibnamefont {Rychetsky}}, \
  and\ \bibinfo {author} {\bibfnamefont {J.}~\bibnamefont {Hlinka}},\ }\href
  {\doibase 10.1103/PhysRevB.81.144125} {\bibfield  {journal} {\bibinfo
  {journal} {Phys. Rev. B}\ }\textbf {\bibinfo {volume} {81}},\ \bibinfo
  {pages} {144125} (\bibinfo {year} {2010})}\BibitemShut {NoStop}%
\bibitem [{\citenamefont {Morozovska}\ \emph {et~al.}(2012)\citenamefont
  {Morozovska}, \citenamefont {Vasudevan}, \citenamefont {Maksymovych},
  \citenamefont {Kalinin},\ and\ \citenamefont
  {Eliseev}}]{morozovska:2012:anisotropic}%
  \BibitemOpen
  \bibfield  {author} {\bibinfo {author} {\bibfnamefont {A.~N.}\ \bibnamefont
  {Morozovska}}, \bibinfo {author} {\bibfnamefont {R.~K.}\ \bibnamefont
  {Vasudevan}}, \bibinfo {author} {\bibfnamefont {P.}~\bibnamefont
  {Maksymovych}}, \bibinfo {author} {\bibfnamefont {S.~V.}\ \bibnamefont
  {Kalinin}}, \ and\ \bibinfo {author} {\bibfnamefont {E.~A.}\ \bibnamefont
  {Eliseev}},\ }\href {\doibase 10.1103/PhysRevB.86.085315} {\bibfield
  {journal} {\bibinfo  {journal} {Phys. Rev. B}\ }\textbf {\bibinfo {volume}
  {86}},\ \bibinfo {pages} {085315} (\bibinfo {year} {2012})}\BibitemShut
  {NoStop}%
\bibitem [{\citenamefont {Fousek}\ and\ \citenamefont
  {Janovec}(1969)}]{fousek:1969:orientation}%
  \BibitemOpen
  \bibfield  {author} {\bibinfo {author} {\bibfnamefont {J.}~\bibnamefont
  {Fousek}}\ and\ \bibinfo {author} {\bibfnamefont {V.}~\bibnamefont
  {Janovec}},\ }\href@noop {} {\bibfield  {journal} {\bibinfo  {journal} {J.
  Appl. Phys.}\ }\textbf {\bibinfo {volume} {40}},\ \bibinfo {pages} {135}
  (\bibinfo {year} {1969})}\BibitemShut {NoStop}%
\bibitem [{\citenamefont {Kresse}\ and\ \citenamefont
  {Furthm{\"u}ller}(1996)}]{kresse:1996:efficiency}%
  \BibitemOpen
  \bibfield  {author} {\bibinfo {author} {\bibfnamefont {G.}~\bibnamefont
  {Kresse}}\ and\ \bibinfo {author} {\bibfnamefont {J.}~\bibnamefont
  {Furthm{\"u}ller}},\ }\href@noop {} {\bibfield  {journal} {\bibinfo
  {journal} {Comput. Mater. Sci.}\ }\textbf {\bibinfo {volume} {6}},\ \bibinfo
  {pages} {15} (\bibinfo {year} {1996})}\BibitemShut {NoStop}%
\bibitem [{\citenamefont {Dudarev}\ \emph {et~al.}(1998)\citenamefont
  {Dudarev}, \citenamefont {Botton}, \citenamefont {Savrasov}, \citenamefont
  {Humphreys},\ and\ \citenamefont {Sutton}}]{dudarev:1998:electron}%
  \BibitemOpen
  \bibfield  {author} {\bibinfo {author} {\bibfnamefont {S.~L.}\ \bibnamefont
  {Dudarev}}, \bibinfo {author} {\bibfnamefont {G.~A.}\ \bibnamefont {Botton}},
  \bibinfo {author} {\bibfnamefont {S.~Y.}\ \bibnamefont {Savrasov}}, \bibinfo
  {author} {\bibfnamefont {C.~J.}\ \bibnamefont {Humphreys}}, \ and\ \bibinfo
  {author} {\bibfnamefont {A.~P.}\ \bibnamefont {Sutton}},\ }\href {\doibase
  10.1103/PhysRevB.57.1505} {\bibfield  {journal} {\bibinfo  {journal} {Phys.
  Rev. B}\ }\textbf {\bibinfo {volume} {57}},\ \bibinfo {pages} {1505}
  (\bibinfo {year} {1998})}\BibitemShut {NoStop}%
\bibitem [{\citenamefont {Jain}\ \emph {et~al.}(2013)\citenamefont {Jain},
  \citenamefont {Ong}, \citenamefont {Hautier}, \citenamefont {Chen},
  \citenamefont {Richards}, \citenamefont {Dacek}, \citenamefont {Cholia},
  \citenamefont {Gunter}, \citenamefont {Skinner}, \citenamefont {Ceder},\ and\
  \citenamefont {Persson}}]{jain:2013:materials}%
  \BibitemOpen
  \bibfield  {author} {\bibinfo {author} {\bibfnamefont {A.}~\bibnamefont
  {Jain}}, \bibinfo {author} {\bibfnamefont {S.~P.}\ \bibnamefont {Ong}},
  \bibinfo {author} {\bibfnamefont {G.}~\bibnamefont {Hautier}}, \bibinfo
  {author} {\bibfnamefont {W.}~\bibnamefont {Chen}}, \bibinfo {author}
  {\bibfnamefont {W.~D.}\ \bibnamefont {Richards}}, \bibinfo {author}
  {\bibfnamefont {S.}~\bibnamefont {Dacek}}, \bibinfo {author} {\bibfnamefont
  {S.}~\bibnamefont {Cholia}}, \bibinfo {author} {\bibfnamefont
  {D.}~\bibnamefont {Gunter}}, \bibinfo {author} {\bibfnamefont
  {D.}~\bibnamefont {Skinner}}, \bibinfo {author} {\bibfnamefont
  {G.}~\bibnamefont {Ceder}}, \ and\ \bibinfo {author} {\bibfnamefont {K.~A.}\
  \bibnamefont {Persson}},\ }\href {\doibase 10.1063/1.4812323} {\bibfield
  {journal} {\bibinfo  {journal} {APL Materials}\ }\textbf {\bibinfo {volume}
  {1}},\ \bibinfo {pages} {011002} (\bibinfo {year} {2013})}\BibitemShut
  {NoStop}%
\bibitem [{\citenamefont {{Mundy, Julia A and Schaab, Jakob and Kumagai, Yu and
  Cano, Andres and Stengel, Massimiliano and Krug, Ingo P and Gottlob, DM and
  Do{\u{g}}anay, Hatice and Holtz, Megan E and Held, Rainer and
  others}}(2017)}]{mundy:2017:functional}%
  \BibitemOpen
  \bibfield  {author} {\bibinfo {author} {\bibnamefont {{Mundy, Julia A and
  Schaab, Jakob and Kumagai, Yu and Cano, Andres and Stengel, Massimiliano and
  Krug, Ingo P and Gottlob, DM and Do{\u{g}}anay, Hatice and Holtz, Megan E and
  Held, Rainer and others}}},\ }\href@noop {} {\bibfield  {journal} {\bibinfo
  {journal} {Nat. Mater.}\ }\textbf {\bibinfo {volume} {16}},\ \bibinfo {pages}
  {622} (\bibinfo {year} {2017})}\BibitemShut {NoStop}%
\bibitem [{\citenamefont {Rahmanizadeh}\ \emph {et~al.}(2014)\citenamefont
  {Rahmanizadeh}, \citenamefont {Wortmann}, \citenamefont {Bihlmayer},\ and\
  \citenamefont {Bl{\"u}gel}}]{rahmanizadeh:2014:charge}%
  \BibitemOpen
  \bibfield  {author} {\bibinfo {author} {\bibfnamefont {K.}~\bibnamefont
  {Rahmanizadeh}}, \bibinfo {author} {\bibfnamefont {D.}~\bibnamefont
  {Wortmann}}, \bibinfo {author} {\bibfnamefont {G.}~\bibnamefont {Bihlmayer}},
  \ and\ \bibinfo {author} {\bibfnamefont {S.}~\bibnamefont {Bl{\"u}gel}},\
  }\href@noop {} {\bibfield  {journal} {\bibinfo  {journal} {Phys. Rev. B}\
  }\textbf {\bibinfo {volume} {90}},\ \bibinfo {pages} {115104} (\bibinfo
  {year} {2014})}\BibitemShut {NoStop}%
\bibitem [{\citenamefont {Sluka}\ \emph {et~al.}(2013)\citenamefont {Sluka},
  \citenamefont {Tagantsev}, \citenamefont {Bednyakov},\ and\ \citenamefont
  {Setter}}]{sluka:2013:free}%
  \BibitemOpen
  \bibfield  {author} {\bibinfo {author} {\bibfnamefont {T.}~\bibnamefont
  {Sluka}}, \bibinfo {author} {\bibfnamefont {A.~K.}\ \bibnamefont
  {Tagantsev}}, \bibinfo {author} {\bibfnamefont {P.}~\bibnamefont
  {Bednyakov}}, \ and\ \bibinfo {author} {\bibfnamefont {N.}~\bibnamefont
  {Setter}},\ }\href@noop {} {\bibfield  {journal} {\bibinfo  {journal} {Nature
  Communications}\ }\textbf {\bibinfo {volume} {4}},\ \bibinfo {pages} {1808}
  (\bibinfo {year} {2013})}\BibitemShut {NoStop}%
\bibitem [{\citenamefont {Bednyakov}\ \emph {et~al.}(2015)\citenamefont
  {Bednyakov}, \citenamefont {Sluka}, \citenamefont {Tagantsev}, \citenamefont
  {Damjanovic},\ and\ \citenamefont {Setter}}]{bednyakov:2015:formation}%
  \BibitemOpen
  \bibfield  {author} {\bibinfo {author} {\bibfnamefont {P.~S.}\ \bibnamefont
  {Bednyakov}}, \bibinfo {author} {\bibfnamefont {T.}~\bibnamefont {Sluka}},
  \bibinfo {author} {\bibfnamefont {A.~K.}\ \bibnamefont {Tagantsev}}, \bibinfo
  {author} {\bibfnamefont {D.}~\bibnamefont {Damjanovic}}, \ and\ \bibinfo
  {author} {\bibfnamefont {N.}~\bibnamefont {Setter}},\ }\href@noop {}
  {\bibfield  {journal} {\bibinfo  {journal} {Scientific Reports}\ }\textbf
  {\bibinfo {volume} {5}},\ \bibinfo {pages} {15819} (\bibinfo {year}
  {2015})}\BibitemShut {NoStop}%
\bibitem [{\citenamefont {Liu}\ \emph {et~al.}(2015)\citenamefont {Liu},
  \citenamefont {Zheng}, \citenamefont {Koocher}, \citenamefont {Takenaka},
  \citenamefont {Wang},\ and\ \citenamefont {Rappe}}]{liu:2015:ferroelectric}%
  \BibitemOpen
  \bibfield  {author} {\bibinfo {author} {\bibfnamefont {S.}~\bibnamefont
  {Liu}}, \bibinfo {author} {\bibfnamefont {F.}~\bibnamefont {Zheng}}, \bibinfo
  {author} {\bibfnamefont {N.~Z.}\ \bibnamefont {Koocher}}, \bibinfo {author}
  {\bibfnamefont {H.}~\bibnamefont {Takenaka}}, \bibinfo {author}
  {\bibfnamefont {F.}~\bibnamefont {Wang}}, \ and\ \bibinfo {author}
  {\bibfnamefont {A.~M.}\ \bibnamefont {Rappe}},\ }\href@noop {} {\bibfield
  {journal} {\bibinfo  {journal} {J. Phys. Chem. Lett.}\ }\textbf {\bibinfo
  {volume} {6}},\ \bibinfo {pages} {693} (\bibinfo {year} {2015})}\BibitemShut
  {NoStop}%
\bibitem [{\citenamefont {Liu}\ and\ \citenamefont
  {Cohen}(2017)}]{liu:2017:stable}%
  \BibitemOpen
  \bibfield  {author} {\bibinfo {author} {\bibfnamefont {S.}~\bibnamefont
  {Liu}}\ and\ \bibinfo {author} {\bibfnamefont {R.}~\bibnamefont {Cohen}},\
  }\href@noop {} {\bibfield  {journal} {\bibinfo  {journal} {J. Phys.: Condens.
  Matter}\ }\textbf {\bibinfo {volume} {29}},\ \bibinfo {pages} {244003}
  (\bibinfo {year} {2017})}\BibitemShut {NoStop}%
\bibitem [{\citenamefont {Li}\ \emph {et~al.}(2018)\citenamefont {Li},
  \citenamefont {Tao}, \citenamefont {Velev},\ and\ \citenamefont
  {Tsymbal}}]{li:2018:resonant}%
  \BibitemOpen
  \bibfield  {author} {\bibinfo {author} {\bibfnamefont {M.}~\bibnamefont
  {Li}}, \bibinfo {author} {\bibfnamefont {L.~L.}\ \bibnamefont {Tao}},
  \bibinfo {author} {\bibfnamefont {J.~P.}\ \bibnamefont {Velev}}, \ and\
  \bibinfo {author} {\bibfnamefont {E.~Y.}\ \bibnamefont {Tsymbal}},\
  }\href@noop {} {\bibfield  {journal} {\bibinfo  {journal} {Phys. Rev. B}\
  }\textbf {\bibinfo {volume} {97}},\ \bibinfo {pages} {155121} (\bibinfo
  {year} {2018})}\BibitemShut {NoStop}%
\bibitem [{\citenamefont {Yang}\ \emph {et~al.}(2015)\citenamefont {Yang},
  \citenamefont {Bhatnagar},\ and\ \citenamefont
  {Alexe}}]{yang:2015:electronic}%
  \BibitemOpen
  \bibfield  {author} {\bibinfo {author} {\bibfnamefont {M.}~\bibnamefont
  {Yang}}, \bibinfo {author} {\bibfnamefont {A.}~\bibnamefont {Bhatnagar}}, \
  and\ \bibinfo {author} {\bibfnamefont {M.}~\bibnamefont {Alexe}},\
  }\href@noop {} {\bibfield  {journal} {\bibinfo  {journal} {Advanced
  Electronic Materials}\ }\textbf {\bibinfo {volume} {1}},\ \bibinfo {pages}
  {1500139} (\bibinfo {year} {2015})}\BibitemShut {NoStop}%
\bibitem [{\citenamefont {Clark}\ and\ \citenamefont
  {Robertson}(2009)}]{clark:2009:energy}%
  \BibitemOpen
  \bibfield  {author} {\bibinfo {author} {\bibfnamefont {S.}~\bibnamefont
  {Clark}}\ and\ \bibinfo {author} {\bibfnamefont {J.}~\bibnamefont
  {Robertson}},\ }\href@noop {} {\bibfield  {journal} {\bibinfo  {journal}
  {Appl. Phys. Lett.}\ }\textbf {\bibinfo {volume} {94}},\ \bibinfo {pages}
  {022902} (\bibinfo {year} {2009})}\BibitemShut {NoStop}%
\bibitem [{\citenamefont {Momma}\ and\ \citenamefont
  {Izumi}(2011)}]{momma:2011:vesta}%
  \BibitemOpen
  \bibfield  {author} {\bibinfo {author} {\bibfnamefont {K.}~\bibnamefont
  {Momma}}\ and\ \bibinfo {author} {\bibfnamefont {F.}~\bibnamefont {Izumi}},\
  }\href {\doibase 10.1107/S0021889811038970} {\bibfield  {journal} {\bibinfo
  {journal} {{Journal of Applied Crystallography}}\ }\textbf {\bibinfo {volume}
  {44}},\ \bibinfo {pages} {1272} (\bibinfo {year} {2011})}\BibitemShut
  {NoStop}%
\end{thebibliography}%

\end{document}